\documentclass[5p,twocolumn,sort&compress]{elsarticle}
\usepackage{graphicx}
\usepackage{array}
\usepackage[fleqn]{amsmath}
\usepackage{amssymb}
\usepackage{amsfonts}
\usepackage{enumitem}
\usepackage[english]{babel}
\usepackage[T2A]{fontenc}
\usepackage{xcolor}
\definecolor{ivan}{rgb}{0.6,0,0}

\newcommand\D{\mathrm{d}}
\newcommand\E{\mathrm{e}}

\begin{document}
\title{Modeling and Controlling the Spread of Epidemic with\\Various Social and Economic Scenarios}
\author{S.\,P.~Lukyanets}
\ead{lukyan@iop.kiev.ua}
\author{I.\,S.~Gandzha}
\ead{gandzha@iop.kiev.ua}
\author{O.\,V.~Kliushnychenko}
\ead{kliushnychenko@iop.kiev.ua}
\address{Institute of Physics, National Academy of Sciences of Ukraine, Prosp.~Nauky 46, Kyiv 03028, Ukraine}

\date{\today}
\begin{abstract}
We propose a dynamical model for describing the spread of epidemics. This model is an extension of the SIQR (susceptible-infected-quarantined-recovered) and SIRP (susceptible-infected-recovered-pathogen) models used earlier to describe various scenarios of epidemic spreading. As compared to the basic SIR model, our model takes into account two possible routes of contagion transmission: direct from the infected compartment to the susceptible compartment and indirect via some intermediate medium or fomites. Transmission rates are estimated in terms of average distances between the individuals in selected social environments and characteristic time spans for which the individuals stay in each of these environments. We also introduce a collective economic resource associated with the average amount of money or income per individual to describe the socioeconomic interplay between the spreading process and the resource available to infected individuals. The epidemic-resource coupling is supposed to be of activation type, with the recovery rate governed by the Arrhenius-like law. Our model brings an advantage of building various control strategies to mitigate the effect of epidemic and can be applied, in particular, to modeling the spread of COVID-19.
\end{abstract}

\maketitle

\section{Introduction}

The spread of contagions, deceases, information, rumors, ideas, or concepts share many similarities. In most cases, such spreading processes are governed by similar models \cite{ExplosiveContagion_2016,PRL_2020,Yu,PRE_2019_Choe,Swiss_PRL_2017,IEEE_2016}, which can vary in terms of complexity or approximations. The simplest dynamical models (see, e.g., Refs.~\cite{Bailey1986,SIAM2000,MultiPatch_2015,SEIRD_2020,Kaxiras_2020,BookSpringerHistory,BookSpringer2015,BookSpringer2019}) have their origin in classical works \cite{Bernoulli,Ross,KermackMcKendrick_I,KermackMcKendrick_II,KermackMcKendrick_III}. More complex models include stochastic effects \cite{Wilkinson_2018,Chaos_2020_Stochastic,Wang_2020_NonlDyn}, spatial flows (e.g., diffusion \cite{Diffusion_2007,Kolokolnikov_2020_Diffusion}), or allow for nontrivial spatial structure or topology \cite{RevModPhys_2015,Hasegawa2016,Masuda_2020_PRR,Croccolo_2020,Arenas_2020_PRX,Winter_2021_Nature}. One common feature in the majority of these models is the presence of kinetic coefficients or parameters that characterize the probability of elementary processes (reactions) per unit time.

For instance, contagion, infection, or decease spreading is associated with scattering of infected individuals on the spatiotemporal fluctuations of population density. The probability of being infected is not a monotonous function of population density \cite{hao,Kolokolnikov_2021}. The probability of scattering of infected individuals on susceptible individuals is lower at low densities, which results in a smaller total scattering cross-section. At higher densities, the mobility of spreaders decreases suppressing the spreading process. Moreover, the scattering of spreaders on local population-density fluctuations should determine the critical concentration of secondary infected cases sufficient for the initiation of the collective process, i.e., an epidemic. The kinetic description of spreading is generally quite a complicated problem that should take account of the internal state of spreaders, which changes in the course of collisions, the presence of spatial inhomogeneities, and non-equilibrium properties of the system.

The dynamics of any spreading process such as an epidemic is determined by individual peculiarities of people in a selected social group, by the type and mechanism of infection, etc. It can also be affected by the ways the process is controlled and influenced. A dramatic example is the spread of COVID-19 (SARS-CoV-2 coronavirus), when different countries and governments resorted to different control strategies and quarantine measures~\cite{Anderson_Lancet_2020-03,Sharov_2020}. Such measures can be controlled by the choice of certain parameters accessible to society (e.g., working day duration, average density of people in public places, frequency of disinfection, intensity of transport communications, etc.). The problem of strategy selection reduces to problems of optimal control theory for feedback systems \cite{Economics_2014,Optimization_SIS_2017,Percolation_2017} or to the theory of games in a more general case \cite{Bauch_Earn_2004}.

On the other hand, kinetic coefficients or probabilities are not the proper control parameters that are readily available, and they can only be estimated indirectly in terms of other parameters. For example, the recovery rate is generally determined by the quality of provision with medical services and food, apart from the individual peculiarities of the given member of population (see, e.g., Refs.~\cite{SIRS_2012,Swiss_2015,China_PRE_2019,SIRS_Chaos_2020,SEIR_Chaos_2020,Boccaletti_Chaos_2020}). The quickest recovery depends on the cost of medical services and the bare subsistence level of consumption $E$, as well as on the availability of collective economic resource $\rho$. Since the cost of services is fixed, the service is terminated if there is no sufficient resource ($\rho\ll E$). In other words, the parameter $E$ serves as the height of some energy barrier peculiar to the given system. Therefore, the recovery rate can be supposed to have an activation-type (Arrhenius-like) dependence, $\propto\exp(-E/\rho)$, similar to the temperature dependence of common activation processes with activation energy $E$~\cite{Laidler,Stiller}.

In this case, the economic resource $\rho$ formally plays the role of effective market temperature and the minimum level of resource consumption is associated with activation energy $E$ \cite{ArXiv_Collapse}. This observation follows from the fact the effective market temperature can be associated with the average amount of money or average income per individual (or economic agent) and the equilibrium distribution of income or money for single economic agents is governed by the exponential Boltzmann law (at least for low and middle income classes) \cite{Yakovenko_2000,Yakovenko_2009,Yakovenko_2010,Yakovenko_2019}.

The activation process implies that the system can exhibit the so-called explosive (or catastrophic) instability \cite{PRL_1966}. For example, when a chemical reaction occurs with the release of heat and has an activation character, it goes faster at higher temperatures. This leads to yet greater temperature increase and ultimately to a thermal explosion, which is described in the framework of the Zel'dovich-Frank-Kamenetskii theory \cite{Zeldovich_Frank,Smirnov,Novozh_2018}.

The fight against the epidemic involves similar catastrophic processes. The spread of epidemic and the associated quarantine measures result in the reduction of the collective resource $\rho$. When the resource is depleted, the quality of medical services drops and the recovery rate goes down. As a result, the number of active members in population decreases. This, in turn, leads to a further reduction of the collective resource, with the level of income needed for the basic survival being lower and lower. Such a scenario can finally result in the ultimate collapse of the system---the effect opposite to thermal explosion \cite{ArXiv_Collapse}.

In this paper, we address some of the points mentioned above. In other words, we are interested not only in modeling the spread of epidemic as an example of a spreading process but also in identifying the ways it can be controlled by socially available parameters and which consequences such control measures can lead to. To demonstrate a number of possible effects, we turn to a quite simple dynamical feedback model. This model has the following two important features.

(I) To take into account the fact that an individual can stay in different environments or public places (e.g., transport, shop, work) characterized by different densities of surrounding individuals, it is convenient to consider some social group and its average daily cycle ($T$). For instance, we can introduce the average daily time spans for which the individual stays at home ($T_1$), at shop or other public places ($T_2$), in transport ($T_3$), at work ($T_4$), etc. and define (at least under normal conditions) the characteristic density $c_j$ of individuals or the average distance $\overline{\ell_j}$ between them in each of the locations $j$. We also make a qualitative assessment of the dependence of the transmission rate on the density of individuals or average distance between them in order to clearly identify the set of control parameters $\{T_j,\,\overline{\ell_j}\}$. Finally, we take into consideration a possibility of indirect transmission through the medium (e.g., contaminated water \cite{SIWR2010,SIRPn2011}) or via the so-called fomites~\cite{Fomite2009,Fomite2018}. The indirect transmission rate depends on the same set of control parameters $\{T_j,\,\overline{\ell_j}\}$.

(II) The interplay between the epidemic and the economic resource available for the selected social group is taken into account by means of the activation-type dependence of characteristic recovery rates. To this end, we use the simplest form of the equation for the dynamics of the collective resource~$\rho$.

This paper is organized as follows. In Section 2, we estimate the direct and indirect transmission rates of contagion in terms of basic social control parameters such as the average distance between the individuals and the average time spent in a certain location. In Section~3, we consider an extended suspected-infected-quarantine-recovered (SIR/SIQR) model that takes into consideration the above-discussed points (I) and (II). In Section 4, we provide several examples of numerical modeling. In particular, we demonstrate the effect of indirect transmission, the effect of quarantine measures, and the effect of limited resource. Conclusions are drawn in Section 5.

\section{Direct and indirect transmission rates}

As mentioned above, the spread of contagion such as infection can be controlled only by the parameters accessible to society. For example, the average distance $\overline{\ell_j}$ between the individuals and the average time $T_j$ spent in a certain location $j$ can be such control parameters. Here we estimate the infection transmission rates in terms of these parameters in the case of direct transmission (infected-to-susceptible) and indirect transmission (through the environment).

\begin{table*}[t]
\centering
 \caption[]{\label{tab:omegaj}Social control parameters and transmission rates for $\ell_0 = 1$~m, $\ell_c = 4$~m, and $\tau_{is}=T/3$ ($T=1$ day).}
\footnotesize
\setlength{\extrarowheight}{2pt}
\vspace{3pt}
\begin{tabular}{ll|lll|lll|lll}
\hline
Location & Description & $T_j$, h & $\overline{\ell_j}$, m & $\omega_j$ & $T_j$, h & $\overline{\ell_j}$, m & $\omega_j$ & $T_j$, h & $\overline{\ell_j}$, m & $\omega_j$  \\
\hline
$j=1$  & Home      & 11.5 & 3.5 & $0.06\,T^{-1}$ & 12.5& 3.5  & $0.06\,T^{-1}$ & 19.5 & 3.5  & $0.06\,T^{-1}$ \\
$j=2$  & Shopping  & 1.5  & 1.5 & $1.15\,T^{-1}$ & 1.5 & 2.25 & $0.41\,T^{-1}$ & 0.5  & 2.25 & $0.41\,T^{-1}$ \\
$j=3$  & Transport & 3    & 1   & $2.81\,T^{-1}$ & 2   & 1.5  & $1.15\,T^{-1}$ & 0    &      &     \\
$j=4$  & Work      & 8    & 3   & $0.15\,T^{-1}$ & 8   & 3.25 & $0.10\,T^{-1}$ & 4    & 3.25 & $0.10\,T^{-1}$ \\
\hline
&$\beta$ & \multicolumn{2}{l}{Casual} &$0.5\,T^{-1}$& \multicolumn{2}{l}{Soft quarantine} & $0.18\,T^{-1}$ & \multicolumn{2}{l}{Strict quarantine} & $0.07\,T^{-1}$\\
\end{tabular}\vspace{3pt}
\end{table*}

\subsection{Direct transmission}
It can naturally be assumed that the probability of being infected, which generally depends on the distance between the infected and susceptible individuals, is described by a damped function with a characteristic correlation radius $\ell_c$ (decay length). For simplicity, we suppose that the susceptible individual and the infected individual can interact only when they fall into the vicinity of radius $\ell_c$ with each other and that the probability of elementary transmission per unit time in this case does not depend on distance and equals $\omega$. We also assume that for any susceptible individual the transmissions from two different infected individuals are independent events. Note that in our simplified consideration the correlation radius $\ell_c$ can be associated with a socially safe distance.

Next we suppose that for each social location $j$ the average population density $\overline{c_j}$ is not affected by any social conditions and is solely determined by the average distance between individuals, $\overline{c_j}\approx (\sqrt{\pi}\overline{\ell_j})^{-2}$. The average number of individuals falling into the circle of radius $\ell_c$ around the susceptible individual in each of the locations $j$ is $\overline{N}_j\approx\overline{c_j}\pi\ell_c^2$. Then the probability for the susceptible individual to be infected over time interval $\Delta t$ can be estimated as $\Delta P_j = \omega \kappa(\overline{N_j}-1)\Delta t$. Here $\kappa\approx i = I/N$ is the probability that the given individual in the vicinity of the susceptible individual is infected, $N$ being the total number of individuals, $I$ being the total number of infected individuals, and $i$ being the number density of infected individuals. Summing over all the susceptible individuals $S$, we obtain an increment of the number of infected individuals, $\Delta I = S\Delta P_j$. Dividing by the total population $N$ and the time interval $\Delta t$, we get a characteristic growth rate of the number density of infected individuals:
\begin{equation}\label{eq:cross-section}
\frac{\Delta i}{\Delta t}\approx \omega (\overline{N_j}-1)is,
\end{equation}
where $s=S/N$ is the number density of susceptible individuals. Thus, the direct (infected-to-susceptible) transmission rate of infection in the selected area with population density $\overline{c_j}$ is estimated as
\begin{equation*}
\omega_j=\omega (\overline{N_j}-1)\approx\omega \left(\left(\frac{\ell_c}{\overline{\ell_j}}\right)^2-1\right).
\end{equation*}

The probability $\omega$ of elementary transmission per unit time can be estimated in terms of the minimum possible distance $\ell_0$ between two individuals (the so-called close contact distance). Assume that the suspected individual is infected with probability equal to unity if he stays at the distance $\ell_0$ from the infected individual for some characteristic time $\tau_{is}$. Then the probability of becoming infected at the minimum distance per unit time is proportional to $\tau_{is}^{-1}$, so that $\omega = a\,\tau_{is}^{-1}$. The parameter $a$ can be estimated as the ratio of the contact area between the susceptible individual and the infected individual to the total area determined by the correlation radius $\ell_c$: $a\approx (\ell_0/\ell_c)^2$. Thus, we have
\begin{equation}\label{eq:omega_j}
\omega_j\approx\tau_{is}^{-1}\left(\frac{\ell_0}{\ell_c}\right)^2 \left(\left(\frac{\ell_c}{\overline{\ell_j}}\right)^2-1\right).
\end{equation}

Let us assume that each individual can stay in $N_\omega$ different locations during some typical period $T$ (e.g., one day). Each location is characterized by its own average distance $\overline{\ell_j}$. Then the integral transmission rate $\beta$ for all locations is given by the following formula:
\begin{equation}\label{eq:beta}
\beta = \sum_{j=1}^{N_\omega}\omega_j\frac{T_j}{T},
\end{equation}
where $T_j$ are the characteristic daily time spans spent in each of the locations $j$, with
\begin{equation}\label{eq:Tj}
\sum_{j=1}^{N_\omega}T_j=T.
\end{equation}

As an example, consider four different social locations ($N_\omega = 4$). Location 1 refers to staying at home (limited social contacts), location 2 refers to shopping and other social contacts during a day, location 3 refers to staying in public transport (where the transmission probability is the highest), location 4 refers to staying at work. Table~\ref{tab:omegaj} gives the transmission rates $\omega_j$ and $\beta$ calculated for various sets of social control parameters $T_j$ and $\overline{\ell_j}$. We considered three possible scenarios: casual (no restrictions), soft quarantine (social distancing in place), and strict quarantine (restricted public transport, limited social contacts, partial cutting of economic activity). The soft quarantine measures reduce the integral transmission rate $\beta$ by a factor of 3, and the strict quarantine measures reduce it by a factor of 7. It can be seen that our rough estimates given by formulas (\ref{eq:omega_j}) and (\ref{eq:beta}) allow the transmission rates to be controlled by the proper choice of parameters $T_j$ and $\overline{\ell_j}$ in different social scenarios. Our $\beta$ estimates in the case of casual scenario fall in the range ($\approx\,$$0.6\,T^{-1}$) derived from the statistical data for the COVID-19 epidemic in the Wuhan city \cite{Covid19_Wuhan_Elsevier}.

Note that the $\omega_j$ estimate given by formula (\ref{eq:omega_j}) is very rough. In fact, the transmission rate depends on fluctuations in the number of individuals falling into the circle of radius $\ell_c$. The maximum number $N_m$ of individuals that can fall into the circle of radius $\ell_c$ is determined by the closest packing of the region $r<\ell_c$ by hard spheres with radius equal to the the minimum possible distance $\ell_0$ between two individuals. For example, considering a triangular lattice with lattice constant $\ell_0$, we have
\begin{equation*}
N_m\approx 1+ 6 \left(\frac{\ell_c}{\ell_0}\right)\left(\frac{\ell_c}{\ell_0}+1\right).
\end{equation*}
Let $\vartheta$ be the probability that the lattice node is occupied. Then the average number of individuals falling into the region $r<\ell_c$ is $\overline{N}=\vartheta N_{m}$, and its mean-square fluctuation is estimated as
$$
\overline{\Delta N^2}=\overline{N^2}-\overline{N}^2=\vartheta N_m + \vartheta^2 N_m (N_m-1) \approx \vartheta N_m \approx \overline{N}.
$$
Finally, the relative error introduced by such fluctuations in the $\omega_j$ estimate can therefore be estimated as
\begin{equation*}
\frac{\delta \omega_j}{\omega_j}=\frac{\sqrt{\overline{\Delta N^2_j}}}{\overline{N_j}}\approx
\overline{N_j}^{-\frac{1}{2}}\approx \frac{\overline{\ell_j}}{\ell_c}.
\end{equation*}

It should also be noted that the characteristic scattering cross-section given by relation (\ref{eq:cross-section}) can have a more complex dependence on density, $\sim \overline c^\alpha$ ($1\leq\alpha\leq 3$), and be governed by a nonlinear incidence function other than $is$ \mbox{\cite{hao,Schlogl_1972,Liu_1986,Liu_1987,Derrick_1993,Kolokolnikov_2021}}.

\subsection{Indirect transmission}
In addition to the direct transmission of infection by direct contact of susceptible individuals with infected individuals, the infection caused by a pathogen (e.g., virus) can be transmitted indirectly either through the medium (e.g., contaminated water \cite{SIWR2010,SIRPn2011}) or via intermediate objects (like hands, counters, doorknobs, etc.) often called fomites~\cite{Fomite2009,Fomite2018}. Such an intermediate medium or fomites will further be referred to as ``cloud''. We associate a separate cloud with each social location $j$.

Let $d_j$ be the pathogen density per individual in the cloud $j$. Then a characteristic growth rate of the number density of infected individuals attributed to the indirect transmission of the infection from the cloud $j$ to susceptible individuals can be calculated in the same way as in the case of direct transmission, namely
\begin{equation}\label{eq:cross-section_indirect}
\frac{\Delta i}{\Delta t}\approx \Omega_j d_j s.
\end{equation}
Here the parameter $\Omega_j$ describes the typical (indirect) rate of pathogen transmission from the cloud $j$ to a susceptible individual. The estimates of indirect transmission rates for various sets of control parameters $T_j$ and $\overline{\ell_j}$ are provided in \ref{apex:tables}.

\section{Dynamical model}

\subsection{Mathematical formulation}

\begin{figure*}[t]
\centering\includegraphics[width=14cm]{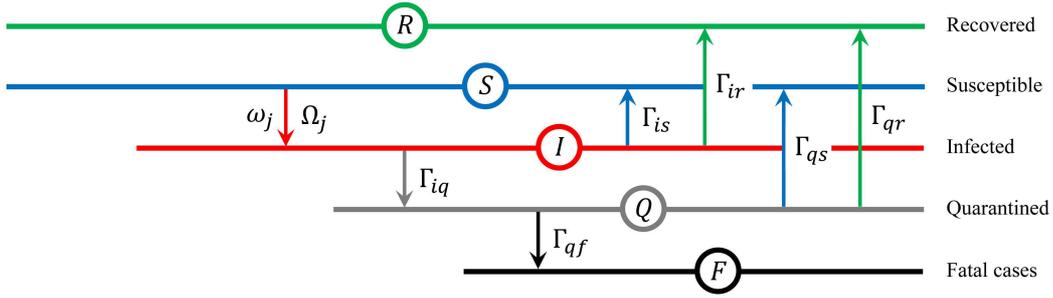}
\caption{\label{fig:scheme}A schematic diagram of the dynamical model given by Eqs.~(\ref{eq:model}) that depicts transitions between different compartments. The estimates of the transition rate constants $\omega_j$, $\Omega_j$, $\Gamma_{is}$, $\Gamma_{ir}$, $\Gamma_{iq}$, $\Gamma_{qs}$, $\Gamma_{qr}$, and $\Gamma_{qf}$ between the compartment levels are given in Table~\ref{tab:coef}.}
\end{figure*}

We subdivide the selected social group into five compartments: susceptible (S), infected (I), quarantined (Q), recovered (R), and deceased (F). Susceptible individuals~(S) are those who are at risk to be infected. Infected individuals (I) are those who have been infected and pass through the virus incubation period. Then they either recover (with or without the acquired immunity) or pass to the severe form of decease (like fever or organ disfunction) when they need to stay isolated either at home or at hospital getting a medical help. Such individuals are referred to as quarantined individuals (Q). The quarantined individuals either recover (with or without the acquired immunity) or die. The corresponding mathematical model is given by the following system of ODEs:
\begin{subequations}\label{eq:model}
\begin{align}
\partial_{t} s &= -s\sum_{j}\Bigl(\omega_j\, i + \Omega_{j}\, d_j\Bigr)\frac{T_j}{T} + \Gamma_{is}\, i + \Gamma_{qs}\,\E^{-E/\rho}\,q,\label{eq:model_siqr_s}\\
\partial_{t} i &=  s\sum_{j}\Bigl(\omega_j\, i + \Omega_{j}\, d_j\Bigr)\frac{T_j}{T}  - \Gamma_i\,i,\label{eq:model_siqr_i}\\
\partial_{t} q &= \Gamma_{iq}\, i - \Gamma_q\,q,\label{eq:model_siqr_q}\\
\partial_{t} r &= \Gamma_{ir}\, i + \Gamma_{qr}\,\E^{-E/\rho}\,q ,\label{eq:model_siqr_r}
\end{align}
where the operator $\partial_t$ stands for the derivative with respect to time $t$. The functions $s(t)$, $i(t)$, $q(t)$, and $r(t)$ describe the number densities of the susceptible, infected, quarantined, and recovered compartments, respectively. Here the index $j=1,\ldots,N_\omega$ runs through all the social locations, and $N_\omega$ is the total number of such locations taken into consideration (see Section 2). The direct transmission rates $\omega_j$ in each of the locations $j$ are defined by formula~(\ref{eq:omega_j}). The characteristic daily time spans $T_j$ in each of the locations $j$ are given in Table~\ref{tab:omegaj}, the total time spent in all the locations being constant each day and equal to the day duration $T$ [see formula~(\ref{eq:Tj})].

Each of the functions $d_j(t)$ describes the pathogen density per individual in the cloud $j$ and contributes to the indirect transmission of the infection from the cloud to susceptible individuals. The pathogen dynamics in the cloud $j$ is given by the following equation:
\begin{equation}\label{eq:model_wj}
\partial_{t} d_j =  \sigma_j\, i - \gamma_j\,d_j.\\
\end{equation}
The first term, $\sigma_j\, i$, describes the pathogen shedding by the infected individuals into the cloud, and the second term, $\gamma_j\, d_j$, describes the pathogen decay in the cloud due to natural inactivation, decontamination, or other routes.

The function $\rho(t)$ represents the average resource associated with the average amount of money or income per individual in the selected social group. The resource balance equation is written as follows \cite{ArXiv_Collapse}
\begin{equation}\label{eq:model_rho}
\partial_{t} \rho = G(s+i+r)-\Gamma_{\rho\,}\rho+\Lambda.
\end{equation}
The acquisition of this resource per unit time is proportional to the number density of working (active) individuals (the quarantined individuals are assumed to be not working). The parameter $G$ formalizes the resource amount acquired by them per unit time. In line with Ref.~\cite{Amado_2019}, we refer to this parameter as acquisition rate. It is proportional to the average working time $T_4$ for the given social group (see Table~\ref{tab:omegaj}). The second term, $\Gamma_\rho\,\rho$, formally describes the collective expenses or taxes. Roughly speaking, the expenses are assumed to be proportional to earnings. Thus, the coefficient $\Gamma_\rho$ represents the resource consumption rate. The parameter $\Lambda$ represents a resource source (constant resource inflow into the system from some external reservoir) or a resource sink (constant resource outflow from the system). When $\Lambda>0$, resource is fed into the system (e.g., in the form of subsidies) from some external source, e.g., a central bank or central government.  When $\Lambda<0$, resource flows out from the system, e.g., in the form of infrastructure expenses, depreciation, rent, interest payments, or other fixed expenses.

The recovery process is governed by the general economic situation characterized by a certain minimum level of resource consumption $E$, which reflects the cost of medical and other essential life services. It defines the minimum amount of the consumed resource needed for the recovery of individuals in the selected social group, therefore implying the existence of some ``energy'' barrier for their recovery. The recovery rates, i.e., the coefficients at the function $q(t)$ in Eqs.~(\ref{eq:model_siqr_s}) and (\ref{eq:model_siqr_r}), are supposed to have an activation-type (Arrhenius-like) dependence, $\propto\exp(-E/\rho)$, similar to the temperature dependence of common activation processes with activation energy $E$.

The number density of fatal cases is described by the function $f(t)$ that is determined by the equation
\begin{equation}\label{eq:model_f}
\partial_{t} f = \left(\Gamma_q +(\Gamma_{qf}-\Gamma_q)\,\E^{-E/\rho}\right) q.
\end{equation}

The above equations are supplemented with the following initial conditions
\begin{equation}
\begin{split}
s(0)&=1-i_0,\;i(0)=i_0,\\
q(0)&=r(0)=f(0)=d_j(0)=0,\;\rho(0)=\rho_0,
\end{split}
\end{equation}
\end{subequations}
$i_0$ being the initial number density of infected individuals and $\rho_0$ being the initial resource value.

The total number density of individuals is assumed to be constant:
\begin{equation}\label{eq:conservation}
s(t)+i(t)+q(t)+r(t)+f(t)=1.
\end{equation}

Effectively, our model is a combination or extension of SIR-like (susceptible-infected-recovered) epidemic models such as SIQR (susceptible-infected-quarantined-recovered) \cite{SIQR1995}, SIWR/SIVR/SIRP (susceptible-infected-recovered-pathogen) \cite{SIWR2010,SIWR2013,SIVR2017,SIRP2018}, SIRD (susceptible-infected-recovered-deceased) \cite{SIRD1,SIRD2}, and EITS (environmental infection transmission system)~\cite{Fomite2009}.

The extension to multiple groups is given in \ref{apex:multigroup}.

\subsection{Assumptions}
\begin{enumerate}[leftmargin=*]
\item The population subsystem is closed (no migration outside the selected population group and no one is added to the population).
\item Natural demography is ignored.
\item A uniform spatial distribution of people is assumed. The pathogen distribution in each cloud is assumed to be uniform.
\item The latent period from exposure to the onset of infectiousness is ignored, i.e., all the exposed individuals are assumed to be infected and can infect others. The SEIR (susceptible-exposed-infected-recovered) model was demonstrated to have no practical advantage as compared to the SIR model~\cite{Covid19_Wuhan_Elsevier}.
\item Quarantined individuals do not infect others.
\item There is no pre-existing immunity in susceptible individuals.
\item There is no loss of immunity by recovered individuals.
\end{enumerate}

\subsection{Parameters}

The full description and indicative values of the parameters of Eqs.~(\ref{eq:model}) are given in Table~\ref{tab:coef} (see \ref{apex:tables}). Figure~\ref{fig:scheme} shows a schematic diagram that depicts transitions between different compartments and identifies the corresponding transition rates.

In particular, the parameter $\Gamma_{is}$ describes a rate at which the infected individuals recover without acquiring the immunity and come back to the susceptible compartment. The parameter $\Gamma_{iq}$ describes a rate at which the infected individuals develop a severe condition and pass to the quarantine compartment, where they become isolated either at home or at hospital. The parameter $\Gamma_{ir}$ describes a rate at which the infected individuals recover without complications and acquire the immunity.  It is proportional to the probability $\mu$ of acquiring the immunity (see Table~\ref{tab:coef}). The rate constant $\Gamma_i$ is defined as follows:
\[
\Gamma_i=\Gamma_{iq}+\Gamma_{ir}+\Gamma_{is} = \tau_i^{-1},
\]
where $\tau_i$ is the characteristic pathogen incubation period.

The parameters $\Gamma_{qr}$ and $\Gamma_{qs}$ describe the rates at which the quarantined individuals recover with or without the acquired immunity. The parameter $\Gamma_{qf}$ describes the fatality rate in the case of unlimited resource ($E\ll\rho$). In this case, the quarantined individuals all get the necessary medical care and the fatalities are only attributed to insuperable health complications (such as concomitant diseases or age factor). In the opposite case, when $E\gg\rho$ (no resource to fight against the epidemic), the fatality rate is at its maximum and is equal to the rate constant
\[
\Gamma_{q}=\Gamma_{qr}+\Gamma_{qs}+\Gamma_{qf} = \tau_q^{-1}
\]
for the quarantined individuals. Here $\tau_q$ is the mean quarantine/hospitalization period. This case refers to the full collapse of the medical system when the quarantined individuals get no medical help or treatment.

Each of the parameters $\Omega_j$ describes the typical rate of pathogen transmission from the cloud $j$ to susceptible individuals. Two other cloud-related parameters, $\sigma_{j}$ and $\gamma_{j}$, are the pathogen shedding rate (infected-to-cloud) and decay rate in the cloud, respectively. Their estimates are provided in \ref{apex:tables}.

Note that the instantaneous number density of quarantined (ill) individuals $q(t)$ is not often a convenient indicator for practical applications. The integral number density of quarantined individuals for a certain period of time can be used instead. It is calculated as follows
\begin{equation}\label{qtotal}
q_{\Sigma} = \Gamma_{iq}\int_{0}^{t}i\,\D \tau.
\end{equation}

In the case of unlimited resource ($E=0$), the integral number density of quarantine individuals in the end of epidemic ($t\rightarrow\infty$) is proportional to the total number density of fatal cases,
\begin{equation}
(q_{\Sigma})_{\infty} = \eta\, f_{\infty},
\end{equation}
where $\eta$ is the probability of the fatal scenario for a quarantined individual.

\subsection{Analysis}
For our further analysis, we first find the stationary solution to Eq.~(\ref{eq:model_wj}) for the cloud $j$:
\begin{equation}
d^*_j=\frac{\sigma_j}{\gamma_j}\, i^*.
\end{equation}
This relation allows us to introduce the dimensionless pathogen concentration in the cloud $j$, namely
\begin{equation}\label{eq:cloud_scaling}
p_j=\frac{\gamma_j}{\sigma_j}\, d_j,
\end{equation}
so that $p^*_j=i^*$. Then Eqs.~(\ref{eq:model}) can be rewritten as
\begin{align}\label{eq:model_siqr_scaled}
\partial_{t} s &= -\Bigl(\beta\, i + \sum_{j}\beta_{j}\, p_j\Bigr) s + \Gamma_{is}\, i + \Gamma_{qs}\,\E^{-E/\rho}\,q,\notag\\
\partial_{t} i &=  \Bigl(\beta\, i + \sum_{j}\beta_{j}\, p_j\Bigr) s  - \Gamma_i\,i,\\
\partial_{t} q &= \Gamma_{iq}\, i - \Gamma_q\,q,\notag\\
\partial_{t} r &= \Gamma_{ir}\, i + \Gamma_{qr}\,\E^{-E/\rho}\,q ,\notag\\
\partial_{t} p_j &=  \gamma_j\left(i - p_j\right).\notag
\end{align}
Here $\beta$ is the direct transmission rate (infected-to-susceptible) given by formula (\ref{eq:beta}). The parameters
\begin{equation}\label{eq:beta_j}
\beta_j=\nu_j \frac{T_j}{T}
\end{equation}
are defined in terms of the scaled indirect transmission rates (infected-cloud-susceptible)
\begin{equation}\label{eq:nu_j}
\nu_j=\Omega_j\frac{\sigma_j}{\gamma_j}
\end{equation}
via each of the clouds $j$.

\begin{figure*}[t]
\includegraphics[width=\textwidth]{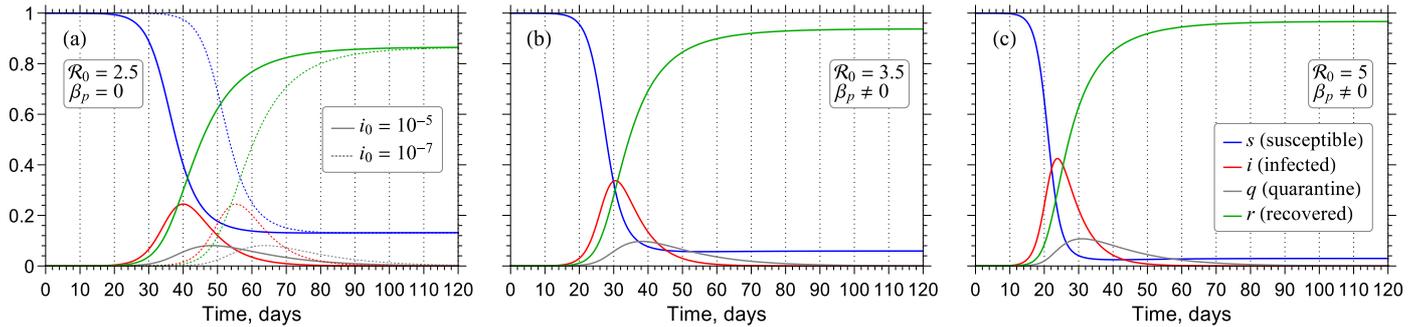}
\caption{\label{fig:cloud}Effect of indirect transmission for fixed $\beta = 0.5\,T^{-1}$ ($T=1$ day). (a) No cloud: $\beta_p=0$, $i_0=10^{-7}$ and $i_0=10^{-5}$. (b) With cloud: $\beta_p=0.2\,T^{-1}$, $i_0=10^{-5}$. (c) With cloud: $\beta_p=0.5\,T^{-1}$, $i_0=10^{-5}$.}
\end{figure*}

\begin{table*}[!]
\centering
 \caption[]{\label{tab:peaks}Main parameters of numerical solutions shown in Figs.~\ref{fig:cloud} and \ref{fig:quarantine}.}
\footnotesize
\setlength{\extrarowheight}{2pt}
\vspace{3pt}
\begin{tabular}{llllcccclll}
\hline
$\beta$ & $\beta_p$ & $\mathcal{R}_0$ & $i_0$ & $i_{\max}$ & $t_{i_{\max}}$, days & $q_{\max}$ & $t_{q_{\max}}$, days & $(q_{\Sigma})_{\infty}$ & $s_{\infty}$\\
\hline
$0.5\,T^{-1}$  & 0              & 2.5  & $10^{-7}$ & 0.245 & 55.4  & 0.081 & 63.7  & 0.193  & 0.132\\
$0.5\,T^{-1}$  & 0              & 2.5  & $10^{-5}$ & 0.245 & 40.0  & 0.081 & 48.3  & 0.193  & 0.132\\
$0.5\,T^{-1}$  & $0.2\,T^{-1}$  & 3.5  & $10^{-5}$ & 0.337 & 30.5  & 0.097 & 38.2  & 0.210  & 0.059\\
$0.5\,T^{-1}$  & $0.5\,T^{-1}$  & 5    & $10^{-5}$ & 0.425 & 23.8  & 0.107 & 31.0  & 0.216  & 0.030\\
\hline
\multicolumn{3}{c}{Quarantine scenario} & $10^{-5}$ & 0.212 & 73.7  & 0.071 & 81.6  & 0.205  & 0.079\\
\hline
\end{tabular}
\end{table*}

Equations (\ref{eq:model_siqr_scaled}) supplemented with Eqs.~(\ref{eq:model_rho}) and (\ref{eq:model_f}) possess two equilibrium points. The first one is the disease-free equilibrium
\begin{gather}
s^{(0)}=1,\;\rho^{(0)}=\frac{G+\Lambda}{\Gamma_\rho},\\
i^{(0)}=q^{(0)}=r^{(0)}=f^{(0)}=p_j^{(0)}=0.\nonumber
\end{gather}
The second one is the endemic equilibrium
\begin{equation}
s^{*}=\mathcal{R}_0^{-1},\quad i^{*}=q^{*}=p_j^{*}=0,
\end{equation}
where the parameter
\begin{equation}\label{eq:R0_SIR}
\mathcal{R}_0=\tau_i\sum_{j}\left(\omega_j+\nu_{j}\right)\frac{T_j}{T}=
\frac{\beta + \beta_p}{\Gamma_{i}}
\end{equation}
is the basic reproduction number and
\begin{equation}\label{eq:beta_p}
\beta_p = \sum_{j}\beta_j
\end{equation}
is the integral indirect pathogen transmission rate via all the clouds.

In general, the basic reproduction number $\mathcal{R}_0$ defines the average number of transmissions one infected individual makes in the entire susceptible compartment during the entire time of being infected. When $\mathcal{R}_0\leqslant 1$, the disease-free equilibrium is stable, and there is no epidemic outbreak. When $\mathcal{R}_0 > 1$, the disease-free equilibrium is unstable, and the system evolves to the state of endemic equilibrium.


The first two equations of system (\ref{eq:model_siqr_scaled}) are strongly nonlinear. There are no analytical solutions known for the general form of these equations. However, in one particular case, when $\Gamma_{is}= \Gamma_{qs}=0$ ($\mu=1$, no loss of immunity) and $E=0$ (unlimited resource), one can get the following asymptotics at $t\rightarrow\infty$:
\begin{equation}\label{eq:sinf}
\log\left(\frac{s(0)}{s_{\infty}}\right)=\mathcal{R}_0\left(1-s_{\infty}\right),\;i_{\infty}=0,\;q_{\infty}=0.
\end{equation}

Equation (\ref{eq:sinf}) can be used to control the accuracy of the numerical integration of system (\ref{eq:model_siqr_scaled}). In the examples considered in the next Section, we used the fourth-order Runge-Kutta method to integrate Eqs.~(\ref{eq:model_siqr_scaled}) numerically with step $\Delta t = T/10$ sufficient to achieve the reasonable accuracy. In the case of $\mu =1$ (no loss of immunity), our numerical estimate of $s_{\infty}$ coincided with the value given by Eq.~(\ref{eq:sinf}) to an accuracy of $10^{-10}$.

\section{Examples}
Now we consider particular examples to demonstrate various effects described by our model. First we focus on the case when there is no resource depletion ($\rho\gg E$), so that the resource activation barrier could be ignored ($E=0$). We illustrate the ``patient zero'' phenomenon, demonstrate the effect of indirect transmission, and model a quarantine scenario. Next we consider an example of a social group with limited resource ($E\ne 0$).

\subsection{Effect of $i_0$}

\begin{figure}[t]
\centering\includegraphics[width=0.9\columnwidth]{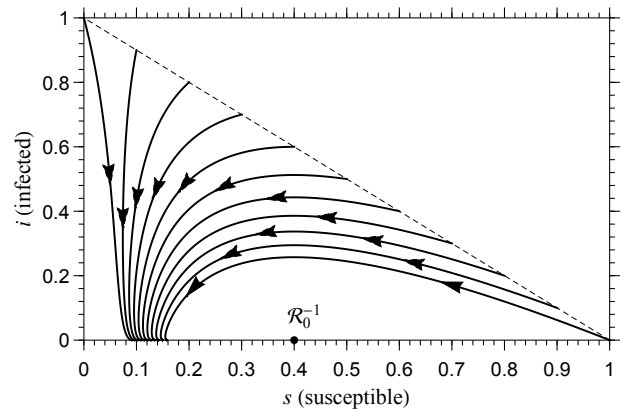}
\caption{\label{fig:plot_si}Susceptible-infected phase plane portrait for $\mathcal{R}_0=2.5$ and various initial number densities of infected individuals~$i_0$.}
\end{figure}

Figure \ref{fig:cloud}a shows the number densities of susceptible, infected, quarantined, and recovered individuals as functions of time in the case of two different initial number densities of infected individuals $i_0$ for the fixed basic reproduction number ($\mathcal{R}_0=2.5$). The case $i_0=10^{-7}$ corresponds to an initial density of one per 10 million, and the case $i_0=10^{-5}$ corresponds to an initial density of one per 100 thousand. The number density of infected individuals exhibits a typical peak and then drops. The peak has the same height for the both initial densities, but in the case of larger $i_0$ it is reached much faster (see Table~\ref{tab:peaks}). This example serves as an illustration of the ``patient zero'' phenomenon. When $\mathcal{R}_0>1$, the epidemic spreads even when it starts only from one infected individual (patient zero). Then it reaches the same intensity in a certain period of time, which is shorter when the initial number of infected individuals is larger. This effect is also clearly seen in the phase portraits $i(s)$ at different $i_0$ (Fig. \ref{fig:plot_si}).

\subsection{Effect of indirect transmission}
Figure \ref{fig:cloud} shows the number densities of susceptible, infected, quarantine, and recovered individuals in the cases when there is no indirect transmission [panel (a)] and when there is such a transmission [panels (b,c)]. The model parameters were selected according to Table~\ref{tab:coef} (see \ref{apex:tables}). The inclusion of the cloud increased the basic reproduction number $\mathcal{R}_0$, so that the peaks of infected ($i_{\max}$) and quarantine ($q_{\max}$) densities might become larger and shift to shorter times (Table~\ref{tab:peaks}).

\subsection{Quarantine scenario}

The epidemic dynamics is governed by the basic reproduction number $\mathcal{R}_0$. The epidemic starts to spread when $\mathcal{R}_0 > 1$. The greater $\mathcal{R}_0$, the larger are $i_{\max}$, $q_{\max}$, and $(q_{\Sigma})_{\infty}$. Thus, to reduce the epidemic peak and to slow the epidemic down, one should reduce $\mathcal{R}_0$. According to formula (\ref{eq:R0_SIR}), this can be achieved by reducing the transmission rates $\omega_j$ and, therefore, the integral transmission rate $\beta$. Such measures are usually referred to as quarantine. To model the quarantine scenario, we assumed the following form of the transmission rates:
\begin{equation}\label{eq:quarantine}
\beta=\left\{\begin{array}{ll}\beta(0),&t<t_1,\\\beta',&t_1\leqslant t<t_2,\\\beta(0),&t\geqslant t_2,\end{array}\right.
\beta_p=\left\{\begin{array}{ll}\beta_p(0),&t<t_1,\\\beta'_{p},&t_1\leqslant t<t_2,\\\beta_p(0),&t\geqslant t_2,\end{array}\right.
\end{equation}
where $t_1$ is the moment when the quarantine starts, $t_2$ is the moment when the quarantine ends, $\beta(0)$ and $\beta_p(0)$ are the direct and indirect transmission rates during the period when there is no quarantine, and $\beta'$ and $\beta'_{p}$ are the transmission rates during the quarantine period.

\begin{figure}[t]
\includegraphics[width=\columnwidth]{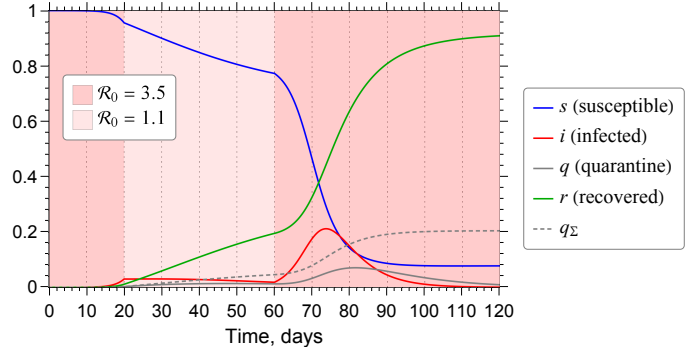}
\caption{\label{fig:quarantine}Number densities of susceptible, infected, quarantined, and recovered individuals in the case of quarantine scenario with $t_1 = 20\,T$, $t_2 = 60\,T$, $\beta(0) = 0.5\,T^{-1}$, $\beta_{p}(0) = 0.2\,T^{-1}$ ($\mathcal{R}_0=3.5$) and $\beta' = 0.18\,T^{-1}$, $\beta'_{p} = 0.04\,T^{-1}$ ($\mathcal{R}_0=1.1$).}
\end{figure}

Figure~\ref{fig:quarantine} demonstrates an example of the quarantine scenario when the initial basic reproduction number $\mathcal{R}_0 = 3.5$ was reduced to $\mathcal{R}_0 = 1.1$ by quarantine measures at the moment $t_1=20$ days. In particular, this can be achieved by increasing the average distance $\overline{\ell_j}$ between the individuals in transport and other social locations and by reducing the average time $T_j$ spent in these locations (see Table~\ref{tab:omegaj} in Sect.~2). The number density of quarantine individuals continued to grow during the quarantine period but at much lesser rate and acquired a local peak at $t\approx 48$~days. Then the quarantine was terminated at the moment $t_2 = 60$ days. The number densities of the infected and quarantined individuals immediately started to grow again and reached the new peaks that were much larger than those during the quarantine (row 5 in Table~\ref{tab:peaks}). As compared to the ``no quarantine'' scenario (Fig.~\ref{fig:cloud}b, row~3 in Table~\ref{tab:peaks}), the absolute heights of the peaks decreased, but the integral number densities of quarantine individuals $q_{\Sigma}$ and, therefore, fatal cases remained nearly the same. Thus, the quarantine scenario allows one to win time but does not seriously affect the total number of ill and deceased people by the end of epidemic, in the case when the mortality rate remains to be constant.

The second (post-quarantine) peak in the number densities of infected and quarantined individuals clearly illustrates the effect known as the second wave of the epidemic, which has in particular been observed in the case of COVID-19 epidemic in many countries \cite{SciRep_2020_secondwave}.

Our results are in line with the results of modeling presented in Ref.~\cite{Anderson_Lancet_2020-03}. The greater the reduction in transmission, the longer and flatter is the epidemic curve, with the risk of resurgence when interventions are lifted to mitigate economic impact. The similar results were obtained when modeling the COVID-19 quarantine scenario for the Wuhan city, with a stochastic SEIR model fitted to the available statistical data \cite{Kucharski_Lancet_2020}. The pre-quarantine $\mathcal{R}_0$ value equal to 2.35 (the median estimate) dropped to $\mathcal{R}_0\approx 1.05$ after the start of the quarantine.

\begin{figure}[t]
\centering\includegraphics[width=0.8\columnwidth]{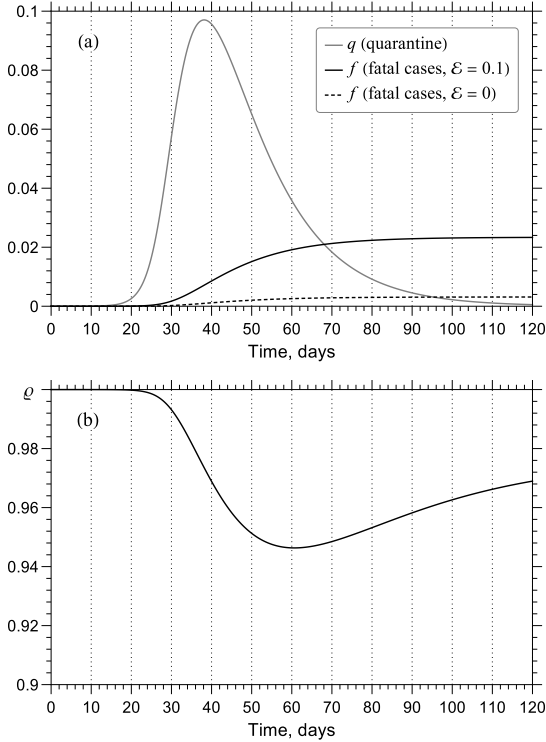}
\caption{\label{fig:resource}Effect of limited resource $\varrho=\rho/\rho^{(0)}$ on the number density of fatal cases $f$ in the case of $\mathcal{E}= E/\rho^{(0)}=0.1$. The number density of quarantined individuals $q$ is the same in the cases of limited ($\mathcal{E}=0.1$) and unlimited ($\mathcal{E}=0$) resource. The model parameters are the same as in the example shown in Fig.~\ref{fig:cloud}b.}
\end{figure}

\begin{figure}[t]
\centering\includegraphics[width=0.85\columnwidth]{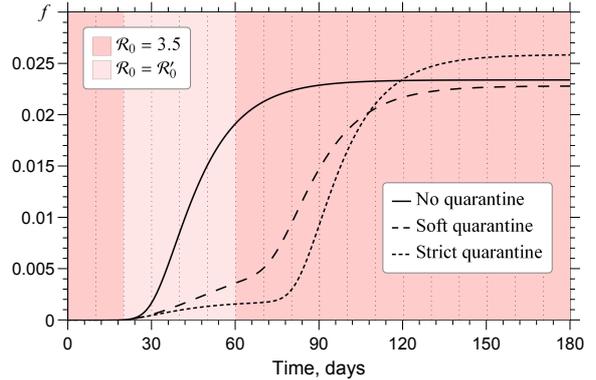}
\caption{\label{fig:resourceG}Effect of two different quarantine scenarios on the number density of fatal cases $f$ in the case of limited resource ($\mathcal{E}=0.1$). The model parameters for the quarantine scenarios (see Table~\ref{tab:omegaj}) with $t_1 = 20\,T$ and $t_2 = 60\,T$ are $\beta(0) = 0.5\,T^{-1}$, $\beta_p(0) = 0.2\,T^{-1}$ (no quarantine, $\mathcal{R}_0=3.5$); $\beta' = 0.18\,T^{-1}$, $\beta'_{p} = 0.04\,T^{-1}$ (soft quarantine, $\mathcal{R}'_0=1.1$); $\beta' = 0.07\,T^{-1}$, $\beta'_{p} = 0.02\,T^{-1}$ (strict quarantine, $\mathcal{R}'_0=0.45$).}
\end{figure}

\subsection{Effect of limited resource}

Here we demonstrate the effect of nonzero resource activation parameter $E$. Let us rewrite resource equation~(\ref{eq:model_rho}) in terms of dimensionless resource $\varrho = \rho/\rho^{(0)}$,
\begin{equation}\label{eq:varrho}
\partial_t \varrho = \frac{\Gamma_\rho}{1+s_\Lambda}\bigl(k(s+i+r)-(1+s_\Lambda)\,\varrho+s_\Lambda\bigr),
\end{equation}
where $\rho^{(0)}=(G^{(0)}+\Lambda)/\Gamma_\rho$ is the equilibrium resource value when there is no epidemic (so that $\varrho^{(0)}\equiv1$), $s_\Lambda=\Lambda/G^{(0)}$ is the number density of active individuals that would have to be working to acquire the resource amount equal to $|\Lambda|$ (per unit time), $k=G/G^{(0)}\leqslant1$ is the time-dependent coefficient that allows for variations in the resource acquisition rate during the spread of the epidemic, and $G^{(0)}$ is the resource acquisition rate when there is no epidemic and the system is in equilibrium. For the sake of simplicity, we will limit our consideration to the case $s_\Lambda=0$.

Resource equation (\ref{eq:varrho}) is coupled to Eqs.~(\ref{eq:model_siqr_scaled}), where the exponential factor $\exp(-E/\rho)$ needs to be identically rewritten as $\exp(-\mathcal{E}/\varrho)$. The parameter $\mathcal{E}\equiv E/\rho^{(0)}$ is the resource activation level $E$ normalized to the equilibrium resource value. Nonzero $\mathcal{E}$ reduces the recovery rates of quarantine individuals, $\Gamma_{qr}\exp(-\mathcal{E}/\varrho)$ and $\Gamma_{qs}\exp(-\mathcal{E}/\varrho)$, so that there would be more fatal cases as compared to the case of unlimited resource ($\mathcal{E}=0$).

Figure~\ref{fig:resource} demonstrates the effect of limited resource in the case of $\mathcal{E}=0.1$, with all other parameters selected such as in the example shown in Fig.~\ref{fig:cloud}b. Nonzero $\mathcal{E}$ has a profound effect on the number of fatal cases, which is nearly 6 times larger than in the case of unlimited resource (Fig.~\ref{fig:resource}a). Such a manyfold increase in the number of fatal cases is caused only by a 5\% drop of resource $\varrho$ (Fig.~\ref{fig:resource}b). Resource began to decline as the number of quarantined individuals grew up (Fig.~\ref{fig:resource}a), since the quarantined individuals are supposed to be passive and not acquiring the resource [see Eq.~(\ref{eq:varrho})]. After the number of quarantined individuals passed through its peak, resource passed through its minimum and started to increase towards its initial value. This example illustrates a scenario when the economic subsystem has a limited capacity to support the medical infrastructure, so that seriously ill (quarantined) individuals could not get the necessary medical help to overcome the infection.

The resource acquisition rate $G$ was assumed to be constant ($k\equiv1$) in the above example. The parameter $G$ is determined by the average number of working hours per working individual and by the working efficiency. The average number of working hours is proportional to the control parameter $T_4$, which can be reduced in the case of quarantine, as discussed in Sect. 2 (see Table~\ref{tab:omegaj}). Figure~\ref{fig:resourceG} demonstrates the effect of two different quarantine scenarios on the number density of fatal cases. When there is no quarantine, the number density of fatal cases is the same as in Fig.~\ref{fig:resource}a for the case of $\mathcal{E}=0.1$. In the case of soft quarantine, the transmission rate $\beta$ goes down (Table~\ref{tab:omegaj}), the basic reproduction number $\mathcal{R}_0$ becoming smaller and the epidemic spreading being less intensive. The soft quarantine scenario does not affect the average number of working hours per working individual, and the resource acquisition rate $G$ remains the same as in the case of no quarantine ($k=1$). The number of fatal cases grows much slower during the quarantine, but it gradually goes back towards its value in the case of no quarantine after the quarantine is terminated.

In the case of strict quarantine, the transmission rate $\beta$ is yet smaller (Table~\ref{tab:omegaj}), mainly because this quarantine scenario also affects the total number of working hours, which is twice as small as compared to the case of soft quarantine. As a result, the resource acquisition rate goes down ($k=0.5$) during the quarantine. In the short run, the number of fatal cases in the case of such strict quarantine measures substantially declines as compared to the cases of soft quarantine and no quarantine (Fig.~\ref{fig:resourceG}). However, it starts to rapidly increase after the quarantine is terminated and eventually becomes larger than it was in the case of no quarantine. This example demonstrates that strict quarantine measures that affect the general economic situation may have serious negative social outcomes in the case of limited economic resource.

\section{Conclusion}

We proposed a dynamical model for describing the spread of epidemics. The spreading process within a selected social group is governed by the equations that explicitly take into account the dependence of characteristic transmission rates on the local population density in various social zones. The indirect channel of transmission via an intermediate environment or the so-called fomites was also taken into consideration. A negative feedback between the infected population size and a collective economic resource associated with the average amount of money or income per individual was introduced to describe the socioeconomic interplay. The epidemic spread and the use of quarantine measures was demonstrated to be connected with economic losses, which in turn could aggravate the negative outcomes of the epidemic.

The model presented in this work can be used to model the COVID-19 epidemic for particular social groups and regions. It can also be applied to describing other spreading processes, such as the spread of information, rumors, ideas, or concepts.

\section*{Acknowledgments}
O.K. was partially supported by a grant for research groups of young scientists from the National Academy of Science of Ukraine (Project No. 0120U100155). We thank Prof. B.I. Lev for fruitful discussions.

\appendix
\setcounter{table}{0}
\section{Model parameters (extended)\label{apex:tables}}

\begin{table*}[!]
\centering
 \caption[]{\label{tab:coef}Model parameters.}
\footnotesize
\setlength{\extrarowheight}{2pt}
\vspace{3pt}
\begin{tabular}{llll}
\hline
 & Description & Our model value & Literature data\\
\hline
$\omega_j$      & Infected-to-susceptible transmission rate for location $j$                & see Table~\ref{tab:omegaj} &\\
$\beta$         & Integral direct transmission rate (infected-to-susceptible)               & $0.5\,T^{-1}$ & $0.6\,T^{-1}$ \cite{Covid19_Wuhan_Elsevier}$^\dag$, $0.15\,T^{-1}$  \cite{Covid19_Wuhan_Cell}$^\dag$\\
$\Omega_j$      & Indirect transmission rate (infected-cloud-susceptible) in cloud $j$      & see Eq.~(\ref{eq:nu_j_estimate}) &\\
$\nu_j$         & Scaled indirect transmission rate in cloud $j$                            & see Eq.~(\ref{eq:nu_j_estimate}) &\\
$\beta_j$       & Scaled indirect transmission rate in cloud $j$ corrected for time span $T_j$  & see Table~\ref{tab:betaj} & \\
$\beta_p$       & Integral indirect transmission rate (infected-cloud-susceptible)          & $0.2\,T^{-1}$   &$0.35\,T^{-1}$ \cite{SIWR2013}$^\ddag$ \\
$\chi_j$        & Infected-cloud-susceptible transmission efficiency in cloud $j$           & $5\times10^{-6}$                 & \\
$\sigma_{j}$    & Average pathogen shedding rate in cloud $j$                               & see Eq.~(\ref{eq:sigma_j})       & \\
$\gamma_{j}$    & Average pathogen decay rate in cloud $j$                              & $\tau_p^{-1}$                 & \\
$\Gamma_{js}$   & Average contact rate (pickup) of susceptible individuals with cloud $j$          & $240\,T^{-1}$                 & 24--480\,$T^{-1}$ (fomites) \cite{Fomite2009}\\
$\Gamma_{ij}$   & Average contact rate (shedding) of infected individuals with cloud $j$         & $360\,T^{-1}$          & $360\,T^{-1}$ (coughs) \cite{Atkinson_2008}$^\S$\\
$\Gamma_{iq}$   & Infected-to-quarantine rate constant                           & $(1-\xi)\,\tau_i^{-1}$        &\\
$\Gamma_{ir}$   & Infected-to-recovered rate constant (with immunity)            & $\xi\,\mu\,\tau_i^{-1}$       &\\
$\Gamma_{is}$   & Infected-to-susceptible rate constant (no immunity)              & $\xi\,(1-\mu)\,\tau_i^{-1}$   &\\
$\Gamma_{i}$    & Integral rate constant for infected individuals, $\Gamma_{iq}+\Gamma_{ir}+\Gamma_{is}$  & $\tau_i^{-1}$   &\\
$\Gamma_{qr}$   & Quarantine-to-recovered rate constant (with immunity)          & $\mu\,(1-\eta)\,\tau_q^{-1}$  &\\
$\Gamma_{qs}$   & Quarantine-to-susceptible rate constant (no immunity)            & $(1-\mu)\,(1-\eta)\,\tau_q^{-1}$ &\\
$\Gamma_{qf}$   & Fatality rate in the case of unlimited resource                       & $\eta\,\tau_q^{-1}$ &\\
$\Gamma_{q}$    & Integral rate constant for quarantined individuals, $\Gamma_{qr}+\Gamma_{qs}+\Gamma_{qf}$& $\tau_q^{-1}$   &\\
$\Gamma_{\rho}$ & Resource consumption rate                                             & $\tau_{\rho}^{-1}$            &\\
$T$             & Unit of time                                                          & 1 day                         &\\
$T_j$           & Average time spent in location $j$                                    & see Table~\ref{tab:omegaj}    &\\
$\ell_0$        & Minimum possible distance between two individuals                     & 1 m    &\\
$\ell_c$        & Correlation radius (the maximum transmission distance)                & 4 m    &\\
$\overline{\ell_j}$        & Average distance between individuals in location $j$                  & see Table~\ref{tab:omegaj}    &\\
$\tau_{is}$     & Characteristic time of becoming infected at close contact             & $T/3$                        & \\
$\tau_p$        & Average pathogen decay time outside the host                          & $2\,T$                        & 0.1--14\,$T^*$ \cite{Stability2020}$^\dag$\\
$\tau_i$        & Average pathogen incubation period                                         & $5\,T$                        &
3--10\,$T$ \cite{Chen2020}$^\dag$, $\approx5\,T$ \cite{IncPeriod2020,NewEngland_2020}$^\dag$\\
$\tau_q$        & Average quarantine/hospitalization time                               & $14\,T$                       &
$(12.4\pm5)\,T$ \cite{Covid19_Wuhan_Cell}$^\dag$, 14.5\,$T$ \cite{Imperial_Report8}$^\dag$\\
$\tau_{\rho}$   & Characteristic resource consumption time                              & $30\,T$                       &\\
$\xi$           & Probability for the infected individual to recover without quarantine & 0.8                           & 0.8$^{**}$ \cite{Anderson_Lancet_2020-03}$^\dag$\\
$\mu$           & Probability of acquiring the immunity                                 & 0.9                           &\\
$\eta$          & Probability of the fatal scenario for the quarantined individual       & 0.015                          & 0.014 \cite{Wu_Nature_2020,Verity_Lancet_2020}$^\dag$\\
$E$             & Minimum level of resource consumption (activation energy)             & 0                            &\\
$G$             & Resource acquisition rate                                             & $G(0)$                       &\\
$\Lambda$       & Resource inflow or outflow per unit time                              & 0                             &\\
$i_0$           & Initial number density of infected individuals                        & $10^{-5}$                     &\\
\hline
\end{tabular}\vspace{3pt}
\raggedright
$^\dag$~for~COVID-19 \\
$^\ddag$ for cholera outbreak \\
$^\S$ for pandemic influenza \\
$^*$ depends on medium, ambient temperature, and surface type \\
$^{**}$ 80\% of COVID-19 cases are mild or asymptomatic
\end{table*}

Table~\ref{tab:omegaj} in Sect. 2 gives the direct transmission rates $\omega_j$ and $\beta$ calculated by formulas (\ref{eq:omega_j}) and (\ref{eq:beta}) for various sets of social control parameters. Table~\ref{tab:coef} gives the full description of the model parameters and their estimates used in our computations.

In particular, the cloud-related parameters can be estimated as follows. The typical (indirect) rate $\Omega_j$ of pathogen transmission from the cloud $j$ to a susceptible individual can roughly be estimated as
\begin{equation}\label{eq:Omega_j}
\Omega_j \approx \Gamma_{js}\frac{\upsilon_{c}}{\Delta}\,\theta, \quad [\,\mathrm{time}^{-1}\times\mathrm{mass}^{-1}\times\mathrm{volume}\,]
\end{equation}
where $\Gamma_{js}$ is the average rate the susceptible individual contacts the cloud $j$, $\upsilon_{c}$ is a typical volume of the pathogen transferred from the cloud to the susceptible individual per one contact, $\Delta$ is some characteristic weight of one pathogen specimen that can be interpreted as the minimum portion (``quant'') of the pathogen that can be transferred per one contact, and $\theta$ is the probability the transmission of this quant results in infection. The smaller the pathogen, the smaller is the quant $\Delta$ and the more intensive is the transmission ($\Omega_j$ is higher). The parameters $\upsilon_{c}$, $\Delta$, and $\theta$ are assumed to be independent of the particular cloud.

The pathogen shedding rate $\sigma_{j}$ (infected-to-cloud) can roughly be estimated as
\begin{equation}\label{eq:sigma_j}
\sigma_j \approx \Gamma_{ij}\frac{n\Delta}{V_j}, \quad [\,\mathrm{time}^{-1}\times\mathrm{mass}\times\mathrm{volume}^{-1}\,]
\end{equation}
where $\Gamma_{ij}$ is the average rate the infected individual contacts the cloud $j$ (number of coughs, sneezes, touches, etc. per unit time), $n$ is the typical number of the pathogen quants $\Delta$ transferred by the infected individual to the cloud per one contact, and $V_j$ is the total volume (capacity) of the cloud $j$. The parameter $n$ is assumed to be independent of the particular cloud.

Then the scaled indirect transmission rates $\nu_j$ can be estimated as
\begin{equation}\label{eq:nu_j_estimate}
\nu_j=\Omega_j\frac{\sigma_j}{\gamma_j}\approx\frac{\Gamma_{ij}\,\Gamma_{js}}{\gamma_j}\,\chi_j.
\end{equation}
The dimensionless parameter
\begin{equation}\label{eq:chi_j}
\chi_j=\frac{\upsilon_{c}}{V_j}\,n\,\theta
\end{equation}
defines the transmission efficiency from infected individuals to susceptible individuals through the cloud $j$. Although the parameter $\Delta$ was eliminated by scaling (\ref{eq:cloud_scaling}), the expression for $\chi_j$ still contains the parameters that are hard to estimate from some physical principles. Therefore, this parameter can rather be estimated by fitting the model to some real statistical data. In practice, it is selected by assuming that the indirect and direct routes of transmissions have approximately the same likelihood, i.e. $\nu_j\approx\omega_j$ \cite{SIWR2010}.

Table~\ref{tab:betaj} gives the indirect transmission rates $\nu_j$ and $\beta_j$ calculated by formulas (\ref{eq:nu_j_estimate}) and (\ref{eq:beta_j}) for a particular set of social control parameters corresponding to the casual scenario in Table~\ref{tab:betaj}. The infected-to-cloud contact rates $\Gamma_{ij}$, pathogen decay rates $\gamma_j$, and transmission efficiencies $\chi_j$ are assumed to be the same for each cloud and listed in Table~\ref{tab:coef}. The cloud-to-susceptible contact rate $\Gamma_{js}$ is assumed to be inversely proportional to the squared average distance $\overline{\ell_j}$ between the individuals [as in Eq.~(\ref{eq:omega_j})].
\begin{table}[t]
\centering
 \caption[]{\label{tab:betaj}Scaled indirect (infected-cloud-susceptible) transmission rates $\nu_j$ and $\beta_j$ in the case of the casual epidemic scenario (see Table~\ref{tab:omegaj}). The aggregate indirect transmission rate $\beta_p$ is given by formula (\ref{eq:beta_p}).}
\footnotesize
\setlength{\extrarowheight}{2pt}
\vspace{3pt}
\begin{tabular}{llllll}
\hline
Cloud & $T_j$, h & $\overline{\ell_j}$, m & $\Gamma_{js}$ & $\nu_j$ & $\beta_j$  \\
\hline
$j=1$  & 11.5 & 3.5 & $20\,T^{-1}$  & $0.07\,T^{-1}$ & $0.03\,T^{-1}$   \\
$j=2$  & 1.5  & 1.5 & $107\,T^{-1}$ & $0.38\,T^{-1}$ & $0.02\,T^{-1}$   \\
$j=3$  & 3    & 1   & $240\,T^{-1}$ & $0.86\,T^{-1}$ & $0.11\,T^{-1}$     \\
$j=4$  & 8    & 3   & $27\,T^{-1}$  & $0.10\,T^{-1}$ & $0.03\,T^{-1}$      \\
\hline
$\beta_p$ &   &     &               &                & $0.20\,T^{-1}$\\
\hline
\end{tabular}
\end{table}

Table~\ref{tab:R0} lists some estimates of the basic reproduction number $\mathcal{R}_0$ derived from the statistical data on COVID-19 (literature data).

\begin{table}[t]
\centering
 \caption[]{\label{tab:R0}Available $\mathcal{R}_0$ estimates for COVID-19.}
 \footnotesize
\setlength{\extrarowheight}{2pt}
\vspace{3pt}
\begin{tabular}{llll}
\hline
$\mathcal{R}_0$ & Reference & Data source  \\
\hline
1.5--3.5  & Imai et al. \cite{Imperial_Report3} & Wuhan \\
2.4--4.1  & Read et al. \cite{Read_R0}          & Wuhan \\
2.2--3.6  & Zhao et al. \cite{Zhao_R0}          & Wuhan \\
1.4--3.9  & Li et al. \cite{NewEngland_2020}    & Wuhan \\
2.5--2.9  & Wu et al. \cite{Wu_Lancet_2020}     & Wuhan \\
\hline
\end{tabular}
\end{table}

\section{Extension to multiple groups\label{apex:multigroup}}
Similarly to the basic SIR model \cite{Bailey1986,MultiPatch_2015}, our model can easily be extended to the multigroup formulation, e.g., with subdivision by age. The dynamics of each group $n$ is governed by the following equations:
\begin{align*}\label{eq:model_multiple}
\partial_{t} s^{(n)} = &-\Bigl(\sum_{m}\beta^{(m)}\, i^{(m)} + \sum_{j}\beta_{j}\, p_j\Bigr) s^{(n)} \\
                       &+ \Gamma_{is}^{(n)}\, i^{(n)} + \Gamma_{qs}^{(n)}\,\E^{-\mathcal{E}/\varrho}\,q^{(n)},\\
\partial_{t} i^{(n)} = &\;\Bigl(\sum_{m}\beta^{(m)}\, i^{(m)} + \sum_{j}\beta_{j}\, p_j\Bigr) s^{(n)}  - \Gamma_i^{(n)}\,i^{(n)},\\
\partial_{t} q^{(n)} = &\;\Gamma_{iq}^{(n)}\, i^{(n)} - \Gamma_q^{(n)}\,q^{(n)},\\
\partial_{t} r^{(n)} = &\;\Gamma_{ir}^{(n)}\, i^{(n)} + \Gamma_{qr}^{(n)}\,\E^{-\mathcal{E}/\varrho}\,q^{(n)}, \\
\partial_{t} f^{(n)} = &\;\left(\Gamma_q^{(n)} +(\Gamma_{qf}^{(n)}-\Gamma_q^{(n)})\,\E^{-\mathcal{E}/\varrho}\right) q^{(n)},
\end{align*}
with index $m$ running over all the groups. The direct and indirect transmission rates are defined as
\begin{equation*}
\beta^{(m)} = \sum_{j=1}^{N_\omega}\omega_j^{(m)}\frac{T_j}{T},\quad
\beta_{j} = \Omega_j\frac{\sum_{m}\sigma_j^{(m)}}{\gamma_j} \frac{T_j}{T}.
\end{equation*}
The pathogen dynamics in the cloud $j$ is given by an equation
\begin{equation*}
\partial_{t} p_j =  \gamma_j\left(\frac{\sum_{m}\sigma_j^{(m)}i^{(m)}}{\sum_{m}\sigma_j^{(m)}} - p_j\right).
\end{equation*}
The resource equation is
\begin{equation*}
\partial_t \varrho = \frac{\Gamma_\rho}{1+s_\Lambda}\Bigl(k\sum_{m}(s^{(m)}+i^{(m)}+r^{(m)})-(1+s_\Lambda)\,\varrho+s_\Lambda\Bigr).
\end{equation*}


\begin{thebibliography}{99}
\bibitem{ExplosiveContagion_2016}
G\'{o}mez-Garde\~{n}es J, Lotero L, Taraskin SN, P\'{e}rez-Reche FJ. Explosive contagion in networks. Sci. Rep. 6, 19767 (2016).

\bibitem{IEEE_2016}
Nowzari C, Preciado VM, Pappas GJ. Analysis and control of epidemics. A survey of spreading processes on complex networks. IEEE Control Syst. Mag. 36, 26--46 (2016).

\bibitem{Swiss_PRL_2017}
B\"{o}ttcher L, Nagler J, Herrmann HJ. Critical behaviors in contagion dynamics. Phys. Rev. Lett. 118, 088301 (2017).

\bibitem{PRE_2019_Choe}
Choe B, Lin Y, Lim S, Lui JCS, Jung K. Efficient spread-size approximation of opinion spreading in general social networks. Phys. Rev. E 100, 052311 (2019).

\bibitem{PRL_2020}
Moore S, Rogers T. Predicting the speed of epidemics spreading in networks. Phys. Rev. Lett. 124, 068301 (2020).

\bibitem{Yu}
Yu S, Yu Z, Jiang H, Mei X, Li J. The spread and control of rumors in a multilingual environment. Nonl. Dyn. 100, 2933--2951 (2020).

\bibitem{Bailey1986}
Bailey NTJ. Macro-modelling and prediction of epidemic spread at community level. Math. Model. 7, 689--717 (1986).

\bibitem{SIAM2000}
Hethcote H.W. The mathematics of infectious diseases. SIAM Rev. 42(4), 599--653 (2000).

\bibitem{MultiPatch_2015}
Bichara D, Kang Y, Castillo-Chavez C, Horan R, Perrings C. SIS and SIR epidemic models under virtual dispersal. Bull. Math. Biol. 77, 2004--2034 (2015).

\bibitem{SEIRD_2020}
Nda\"{\i}rou F, Area I, Nieto JJ, Torres DFM. Mathematical modeling of COVID-19 transmission dynamics with a case study of Wuhan. Chaos Soliton Fract. 135, 109846 (2020).

\bibitem{Kaxiras_2020}
Kaxiras E, Neofotistos G, Angelaki E. The first 100 days: Modeling the evolution of the COVID-19 pandemic. Chaos Soliton Fract. 138, 110114 (2020).

\bibitem{BookSpringerHistory}
Baca\"{e}r N. {\it A Short History of Mathematical Population Dynamics} (Springer, London, 2011).

\bibitem{BookSpringer2015}
Martcheva M. {\it An Introduction to Mathematical Epidemeology} (Springer, New York, 2015).

\bibitem{BookSpringer2019}
Brauer F, Castillo-Chavez C, Feng Z. {\it Mathematical Models in Epidemeology} (Springer, New York, 2019).

\bibitem{Bernoulli}
Bernoulli D. Essai d'une nouvelle analyse de la mortalit\'{e} caus\'{e}e par la petite v\'{e}role. Mem. Math.
Phys. Acad. R. Sci. Paris, 1--45 (1766) [English translation entitled ``An attempt at a new analysis of the mortality caused by smallpox and of the advantages of inoculation to prevent it'' in: Bradley L. {\it Smallpox Inoculation: An Eighteenth Century Mathematical Controversy} (Adult Education Department, Nottingham, 1971), p. 21. Reprinted in: Haberman S, Sibbett TA (Eds.) {\it History of Actuarial Science. Vol. VIII: Multiple Decrement and Multiple State Models} (William Pickering, London, 1995), p. 1.; Blower S. Rev. Med. Virol. 14, 275--288 (2004)]; d'Alembert J. Sur l'application du calcul des probabilit\'{e}s \`{a} l'inoculation de la petite v\'{e}role. In: {\it Opuscules math\'{e}matiques}, t. 2 (David, Paris, 1761), p. 26--95; Dietz K, Heesterbeek JAP. Daniel Bernoulli's epidemiological model revisited. Math. Biosci. 180, 1--21 (2002).

\bibitem{Ross}
Ross R. {\it The Prevention of Malaria}, 2nd edn. (John Murray, London, 1911).

\bibitem{KermackMcKendrick_I}
Kermack WO, McKendrick AG. A contribution to the mathematical theory of epidemics. Proc. Roy. Soc. Lond. A 115, 700--721 (1927).

\bibitem{KermackMcKendrick_II}
Kermack WO, McKendrick AG. Contributions to the mathematical theory of epidemics. II.---The problem of endemicity. Proc. Roy. Soc. Lond. A 138, 55--83 (1932).

\bibitem{KermackMcKendrick_III}
Kermack WO, McKendrick AG. Contributions to the mathematical theory of epidemics. III.---Further studies of the problem of endemicity. Proc. Roy. Soc. Lond. A 141, 94--112 (1933).

\bibitem{Wilkinson_2018}
Wilkinson RR, Sharkey KJ. Impact of the infectious period on epidemics. Phys. Rev. E 97, 052403 (2018).

\bibitem{Chaos_2020_Stochastic}
Bekiros S, Kouloumpou D. SBDiEM: A new mathematical model of infectious disease dynamics. Chaos Soliton Fract. 136, 109828 (2020).

\bibitem{Wang_2020_NonlDyn}
Wang Z, Broccardo M, Mignan A, Sornette D. The dynamics of entropy in the COVID-19 outbreaks. Nonl. Dyn. 101, 1847--1869 (2020).

\bibitem{Diffusion_2007}
Colizza V, Pastor-Satorras R, Vespignani A. Reaction-diffusion processes and metapopulation models in heterogeneous networks. Nat. Phys. 3, 276--282 (2007).

\bibitem{Kolokolnikov_2020_Diffusion}
Gai C, Iron D, Kolokolnikov T. Localized outbreaks in an S-I-R model with diffusion. J.~Math. Biol. 80, 1389--1411 (2020).

\bibitem{RevModPhys_2015}
Pastor-Satorras R, Castellano C, Mieghem PV, Vespignani A. Epidemic processes in complex networks. Rev. Mod. Phys. 87, 925--979 (2015).

\bibitem{Hasegawa2016}
Hasegawa T, Nemoto K. Outbreaks in susceptible-infected-removed epidemics with multiple seeds. Phys. Rev. E. 93, 032324 (2016).

\bibitem{Masuda_2020_PRR}
Masuda N, Holme P. Small inter-event times govern epidemic spreading on networks. Phys. Rev. Res. 2, 023163 (2020).

\bibitem{Croccolo_2020}
Croccolo F, Roman HE. Spreading of infections on random graphs: A percolation-type model for COVID-19. Chaos Soliton Fract. 139, 110077 (2020).

\bibitem{Arenas_2020_PRX}
Arenas A, Cota W, G\'{o}mez-Garde\~{n}es J, G\'{o}mez S, Granell C, Matamalas JT, Soriano-Pa\~{n}os D, Steinegger B. Modeling the spatiotemporal epidemic spreading of COVID-19 and the impact of mobility and social distancing interventions. Phys. Rev. X 10, 041055 (2020).

\bibitem{Winter_2021_Nature}
Wintermantel TM, Buchhold M, Shevate S, Morgado M, Wang Y, Lochead G, Diehl S, Whitlock S. Epidemic growth and Griffiths effects on an
emergent network of excited atoms. Nat. Commun. 12, 103 (2021).

\bibitem{hao}
Hu H, Nigmatulina K, Eckhoff P. The scaling of contact rates with population density for the infectious disease models. Math. Biosci. 244, 125--134 (2013).

\bibitem{Kolokolnikov_2021}
Kolokolnikov T, Iron D. Law of mass action and saturation in SIR model with application to Coronavirus modelling. Infect. Dis. Model. 6, 91--97 (2021).

\bibitem{Anderson_Lancet_2020-03}
Anderson RM, Heesterbeek H, Klinkenberg D, Hollingsworth TD. How will country-based mitigation measures influence the
course of the COVID-19 epidemic? Lancet 395, 931--934 (2020).

\bibitem{Sharov_2020}
Sharov KS. Creating and applying SIR modified compartmental model for calculation of COVID-19 lockdown efficiency. Chaos Soliton Fract. 141, 110295 (2020).

\bibitem{Economics_2014}
Perrings C, Castillo-Chavez C, Chowell G, Daszak P, Fenichel EP, Finnoff D, Horan RD, Kilpatrick AM, Kinzig AP, Kuminoff NV,
Levin S, Morin B, Smith KF, Springborn M. Merging economics and epidemiology to improve the prediction and management of infectious disease. EcoHealth 11, 464--475 (2014).

\bibitem{Optimization_SIS_2017}
Chen H, Li G, Zhang H, Hou Z. Optimal allocation of resources for suppressing epidemic spreading on networks. Phys. Rev. E 96, 012321 (2017).

\bibitem{Percolation_2017}
Schr\"{o}der M, Ara\'{u}jo NAM, Sornette D, Nagler J. Controlling percolation with limited resources. Phys. Rev. E 96, 062302 (2017).

\bibitem{Bauch_Earn_2004}
Bauch CT, Earn DJD. Vaccination and the theory of games, PNAS 101(36), 13391--13394 (2004).

\bibitem{SIRS_2012}
Zhou L, Fan M. Dynamics of an SIR epidemic model with limited medical resources revisited. Nonlinear Anal. Real World Appl. 13, 312--324 (2012).

\bibitem{Swiss_2015}
B\"{o}ttcher L, Woolley-Meza O, Ara\'{u}jo NAM, Herrmann HJ, Helbing D. Disease-induced resource constraints can trigger explosive epidemics. Sci. Rep. 5, 16571 (2015).

\bibitem{China_PRE_2019}
Chen X, Zhou T, Feng L, Liang J, Liljeros F, Havlin S, Hu Y. Nontrivial resource requirement in the early stage for containment of epidemics. Phys. Rev. E 100, 032310 (2019).

\bibitem{SIRS_Chaos_2020}
Mohd MH, Sulayman F. Unravelling the myths of $R_0$ in controlling the dynamics of COVID-19 outbreak: A modelling perspective. Chaos Soliton Fract. 138, 109943 (2020).

\bibitem{SEIR_Chaos_2020}
\c{C}akan S. Dynamic analysis of a mathematical model with health care capacity for COVID-19 pandemic. Chaos Soliton Fract. 139, 110033 (2020).

\bibitem{Boccaletti_Chaos_2020}
Zhang X, Ruan Z, Zheng M, Barzel B, Boccaletti S. Epidemic spreading under infection-reduced-recovery. Chaos Soliton Fract. 140, 110130 (2020).

\bibitem{Laidler}
Glasstone S, Laidler KJ, Eyring H. \textit{The Theory of Rate Processes: The Kinetics of Chemical Reactions, Viscosity, Diffusion and Electrochemical Phenomena} (McGraw-Hill, New York, 1941); Laidler KJ. \textit{Chemical Kinetics}, 3rd ed. (Pearson, 1987).

\bibitem{Stiller}
Stiller W. \textit{Arrhenius Equation and Non-Equilibrium Kinetics: 100 Years Arrhenius Equation} (B.G. Teubner, Leipzig, 1989).

\bibitem{ArXiv_Collapse}
Gandzha IS, Kliushnichenko OV, Lukyanets SP. Epidemic-driven collapse in a system with limited economic resource. Preprint, arXiv:2012.12113 (2020).

\bibitem{Yakovenko_2000}
Dr\u{a}gulescu A, Yakovenko VM. Statistical mechanics of money. Eur. Phys. J. B 17, 723--729 (2000); Evidence for the exponential distribution of income in the USA. Eur. Phys. J. B 20, 585--589 (2001).

\bibitem{Yakovenko_2009}
Yakovenko VM, Rosser Jr JB. Colloquium: Statistical mechanics of money, wealth, and income. Rev. Mod. Phys. 81(4), 1703--1725 (2009).

\bibitem{Yakovenko_2010}
Banerjee A, Yakovenko VM. Universal patterns of inequality. New J. Phys. 12, 075032 (2010).

\bibitem{Yakovenko_2019}
Tao Y, Wu X, Zhou T, Yan W, Huang Y, Yu H, Mondal B, Yakovenko VM. Exponential structure of income inequality: evidence from 67 countries. J. Econ. Interact. Coord. 14, 345--376 (2019).

\bibitem{PRL_1966}
Sturrock PA. Explosive and nonexplosive onsets of instability. Phys. Rev. Lett. 16, 270 (1966).

\bibitem{Zeldovich_Frank}
Zel'dovich YaB, Frank-Kamenetskii DA. On the theory of uniform flame propagation. Dokl. Akad. Nauk SSSR 19, 693 (1938); Frank-Kamenetskii D.A. The temperature distribution in a reaction vessel and the stationary theory of thermal
explosions. Dokl. Akad. Nauk SSSR 18, 413 (1938).

\bibitem{Smirnov}
Smirnov BM. Energetic processes in macroscopic fractal structures. Sov. Phys. Usp. \textbf{34}, 526--541 (1991).

\bibitem{Novozh_2018}
Novozhilov V. Kinetic effects in thermal explosion with oscillating ambient conditions. Sci. Rep. 8, 4030 (2018).

\bibitem{SIWR2010}
Tien JH, Earn DJD. Multiple transmission pathways and disease dynamics in a waterborne pathogen model. Bull. Math. Biol. 72, 1506--1533 (2010).

\bibitem{SIRPn2011}
Shuai Z, van den Driessche P. Global dynamics of cholera models with differential infectivity. Math. Biosci. 234, 118--126 (2011).

\bibitem{Fomite2009}
Li S, Eisenberg JNS, Spicknall IH, Koopman JS. Dynamics and control of infections transmitted from person to person through
the environment. Am. J. Epidemiol. 170(2), 257--265 (2009).

\bibitem{Fomite2018}
Kraay ANM, Hayashi MAL, Hernandez-Ceron N, Spicknall IH, Eisenberg MC, Meza R, Eisenberg JNS. Fomite-mediated transmission as a sufficient pathway: a comparative analysis across three viral pathogens. BMC Infect. Dis. 18, 540 (2018).

\bibitem{Covid19_Wuhan_Elsevier}
Roda WC, Varughese MB, Han D, Li MY. Why is it difficult to accurately predict the COVID-19 epidemic? Infect. Dis. Model. 5, 271--281 (2020).

\bibitem{Schlogl_1972}
Schl\"{o}gl F. Chemical reaction models for non-equilibrium phase transitions. Z. Phys. 253, 147--161 (1972).

\bibitem{Liu_1986}
Liu W, Levin SA, Iwasa Y. Influence of nonlinear incidence rates upon the behavior of SIRS epidemiological models. J.~Math. Biol. 23, 187--204 (1986).

\bibitem{Liu_1987}
Liu W, Hethcote HW, Levin SA. Dynamical behavior of epidemiological models with nonlinear incidence rates. J.~Math. Biol. 25, 359--380 (1987).

\bibitem{Derrick_1993}
Derrick WR, van den Driessche P. A disease transmission model in a nonconstant population. J.~Math. Biol. 31, 495--512 (1993).

\bibitem{Amado_2019}
Amado A, Santana-Filho JV, Campos PRA, Raposo EP. Interplay of sources of stochastic noise in a resource-based model. Eur. Phys. J. Plus 134, 151 (2019).

\bibitem{SIQR1995}
Feng Z, Thieme HR. Recurrent outbreaks of childhood diseases revisited: the impact of isolation. Math. Biosci. 128, 93--130 (1995).

\bibitem{SIWR2013}
Eisenberg MC, Robertson SL, Tien JH. Identifiability and estimation of multiple transmission pathways in cholera and waterborne disease. J. Theor. Biol. 324, 84--102 (2013).

\bibitem{SIVR2017}
Brauer F. A new epidemic model with indirect transmission. J. Biol. Dyn. 11, 285--293 (2017).

\bibitem{SIRP2018}
David JF. Epidemic models with heterogeneous mixing and indirect transmission. J. Biol. Dyn. 12, 375--399 (2018).

\bibitem{SIRD1}
Fanelli D, Piazza F. Analysis and forecast of COVID-19 spreading in China, Italy and France. Chaos Soliton Fract. 134, 109761 (2020).

\bibitem{SIRD2}
Reis RF, Quintela BM, Campos JO, Gomes JM, Rocha BM, Lobosco M, dos Santos RW. Characterization of the COVID-19 pandemic and the impact of uncertainties, mitigation strategies, and underreporting of cases in South Korea, Italy, and Brazil. Chaos Soliton Fract. 136, 109888 (2020).


\bibitem{SciRep_2020_secondwave}
Cacciapaglia G, Cot C, Sannino F. Second wave COVID-19 pandemics in Europe: a temporal playbook. Sci. Rep. 10, 15514 (2020).

\bibitem{Kucharski_Lancet_2020}
Kucharski AJ, Russell TW, Diamond C, Liu Y, Edmunds J, Funk S, Eggo RM. Early dynamics of transmission and control of COVID-19: a mathematical modelling study. Lancet Infect. Dis. 20, 553--558 (2020).


\bibitem{Covid19_Wuhan_Cell}
Wang H, Wang Z, Dong Y, Chang R, Xu C, Yu X, Zhang S, Tsamlag L, Shang M, Huang J, Wang Y, Xu G, Shen T, Zhang X, Cai Y. Phase-adjusted estimation of the number of coronavirus disease 2019 cases in Wuhan, China. Cell Discov. 6, 10 (2020).

\bibitem{Atkinson_2008}
Atkinson MP, Wein LM. Quantifying the routes of transmission for pandemic influenzas. Bull Math. Biol. 70, 820--867 (2008).

\bibitem{Stability2020}
Chin AWH, Chu JTS, Perera MRA, Hui KPY, Yen H-L, Chan MCW, Peiris M, Poon LLM. Stability of SARS-CoV-2
in different environmental conditions. Lancet Microb. 1, e10 (2020).

\bibitem{Chen2020}
Chen J. Pathogenicity and transmissibility of 2019-nCoV---A quick overview and comparison with other emerging viruses. Microb. Infect. 22, 69--71 (2020).

\bibitem{IncPeriod2020}
Lauer SA, Grantz KH, Bi Q, Jones FK, Zheng Q, Meredith HR, Azman AS, Reich NG, Lessler J. The incubation period of coronavirus disease 2019 (COVID-19) from publicly reported confirmed cases: estimation and application. Ann. Intern. Med. 172(9), 577--583 (2020).

\bibitem{NewEngland_2020}
Li Q, Guan X, Wu P, Wang X, Zhou L, Tong Y, Ren R, Leung KSM, Lau EHY, Wong JY, Xing X, Xiang N, Wu Y, Li C, Chen Q, Li D, Liu T, Zhao J, Liu M, Tu W, Chen C, Jin L, Yang R, Wang Q, Zhou S, Wang R, Liu H, Luo Y, Liu Y, Shao G, Li H, Tao Z, Yang Y, Deng Z, Liu B, Ma Z, Zhang Y, Shi G, Lam TTY, Wu JT, Gao GF, Cowling BJ, Yang B, Leung GM, Feng Z. Early transmission dynamics in Wuhan, China, of novel coronavirus–infected pneumonia. N. Engl. J. Med. 382(13), 1199--1207 (2020).

\bibitem{Imperial_Report8}
Gaythorpe K, Imai N, Cuomo-Dannenburg G, Baguelin M, Bhatia S, Boonyasiri A, Cori A, Cucunub\'{a} Z, Dighe A, Dorigatti I, FitzJohn R, Fu H, Green W, Hamlet A, Hinsley W, Laydon D, Nedjati-Gilani G, Okell L, Riley S, Thompson H, van Elsland S, Volz E, Wang H, Wang Y, Whittaker C, Xi X, Donnelly CA, Ghani A, Ferguson NM. Symptom progression of COVID-19. Report 8 of the Imperial College London COVID-19 Response Team (2020). https://doi.org/10.25561/77344.

\bibitem{Wu_Nature_2020}
Wu JT, Leung K, Bushman M, Kishore N, Niehus R, de Salazar PM, Cowling BJ, Lipsitch M, Leung GM. Estimating clinical severity of COVID-19 from the transmission dynamics in Wuhan, China. Nature Medicine 26, 506--510 (2020).

\bibitem{Verity_Lancet_2020}
Verity R, Okell LC, Dorigatti I, Winskill P, Whittaker C, Imai N, Cuomo-Dannenburg G, Thompson H,
Walker PGT, Fu H, Dighe A, Griffin JT, Baguelin M, Bhatia S, Boonyasiri A, Cori A, Cucunub\'{a} Z,
FitzJohn R, Gaythorpe K, Green W, Hamlet A, Hinsley W, Laydon D, Nedjati-Gilani G, Riley S, van Elsland S,
Volz E, Wang H, Wang Y, Xi X, Donnelly CA, Ghani AC, Ferguson NM. Estimates of the severity of coronavirus disease 2019: a model-based analysis. Lancet Infect. Dis. 20, 669--677 (2020).


\bibitem{Imperial_Report3}
Imai N, Cori A, Dorigatti I, Baguelin M, Donnelly CA, Riley S, Ferguson NM. Transmissibility of 2019-nCoV. Report 3 of the Imperial College London COVID-19 Response Team (2020). https://doi.org/10.25561/77148.

\bibitem{Read_R0}
Read JM, Bridgen JRE, Cummings DAT, Ho A, Jewell CP. Novel coronavirus 2019-nCoV: early estimation of epidemiological parameters and epidemic predictions. MedRxiv preprint (2020). https://doi.org/10.1101/2020.01.23.20018549.

\bibitem{Zhao_R0}
Zhao S, Linc Q, Rand J, Musae SS, Yang G, Wangh W, Loue Y, Gaoi D, Yangj L, Hee D, Wanga MH. Preliminary estimation of the basic reproduction number of novel coronavirus (2019-nCoV) in China, from 2019 to 2020: A data-driven analysis in the early phase of the outbreak. Int. J. Infect. Dis. 92, 214--217 (2020).

\bibitem{Wu_Lancet_2020}
Wu JT, Leung K, Leung GM. Nowcasting and forecasting the potential domestic and international spread of the 2019-nCoV outbreak originating in Wuhan, China: a modelling study. Lancet 395, 689--697 (2020).

\end{thebibliography}
\end{document}